# Gluing GAP to RAS Mutants: A New Approach to an Old Problem in Cancer Drug Development

Ivan Ranđelović[1,2], Kinga Nyíri[3,4], Gergely Koppány[3,4], Marcel Baranyi[5], József Tóvári[1,2], Attila Kigyós[1,2], József Timár[5], Beáta G. Vértessy[3,4*], Vince Grolmusz[6,7*]

[1]National Institute of Oncology, Budapest, Hungary
[2]Kinetolab Ltd., Budapest, Hungary
[3]Laboratory of Genome Metabolism and Repair, Institute of Enzymology, Research Centre for Natural Sciences, Hungarian Research Network, Budapest, 1117, Hungary
[4]Department of Applied Biotechnology and Food Science, BME Budapest University of Technology and Economics, Budapest, 1111, Hungary
[5]2nd Department of Pathology, Semmelweis University, Budapest, Hungary
[6]Mathematical Institute of Eötvös Loránd University,1117 Budapest Hungary
[7]Uratim Ltd., Budapest, 1118, Hungary
*corresponding authors: Vince Grolmusz (grolmusz@pitgroup.org) and Beáta G. Vértessy (vertessy.beata@ttk.hu)

**Abstract**

Mutated genes may lead to cancer development in numerous tissues. While more than 600 cancer-causing genes are known today, some of the most widespread mutations are connected to the RAS gene: RAS mutations are found in approximately 25% of all human tumors. Specifically, KRAS mutations are involved in the three most lethal cancers in U.S.: pancreatic ductal adenocarcinoma, colorectal adenocarcinoma, and lung adenocarcinoma. These cancers are among the most difficult to treat, and they are frequently excluded from chemotherapeutic attacks as hopeless cases. The mutated KRAS proteins have specific 3-dimensional conformations, which perturb functional interaction with the GAP protein on the GAP:RAS complex surface leading to a signaling cascade and uncontrolled cell growth. Here we describe a gluing docking method for finding small molecules that bind to both the GAP and the mutated KRAS molecules. These small molecules glue together the GAP and the mutated KRAS molecules and may serve as new cancer drugs for the most lethal, most difficult-to-treat carcinomas. As a proof of concept, we identify two new, drug-like small molecules with the new method: these compounds specifically inhibit the growth of PANC-1 cell line with KRAS mutation G12D *in vitro* and *in vivo*. Importantly, the two new compounds show significantly lower IC50 and higher specificity against the G12D KRAS mutant as compared to the recently described MRTX-1133 inhibitor against the G12D KRAS mutant.

**Introduction**

More than 600 cancer-causing mutated genes (or oncogenes) are known today [1] (cf also listed the Catalogue of Somatic Mutations In Cancer, https://cancer.sanger.ac.uk/cosmic); from these, numerous entries are connected to the RAS gene, whose mutations are found in approximately 25% of all human tumors [2,3]. Cancers caused by RAS mutations are some of the most difficult to treat, and frequently resist chemotherapeutic attacks despite innovative novel approaches [4-6]. Mutations in the RAS genes, and, consequently, in the RAS proteins, are connected to the three most lethal cancers in U.S.: pancreatic ductal adenocarcinoma, colorectal adenocarcinoma, and lung adenocarcinoma [1]. In humans, there are three RAS isoforms: KRAS, NRAS, and HRAS; among these, the KRAS isoform is the most frequently mutated in cancers (>85%) [7]. Therefore, KRAS is one of the most important targets of drug development.

The oncogenic potential of several mutant Ras proteins is directly related to the perturbation of the Ras-GAP interaction. The wild-type RAS molecule binds the GAP (GTPase-activating protein), promotes its GTP hydrolyzing activity, and efficiently shifts the Ras-GTP into Ras-GDP, thereby switching the Ras conformation and stopping signaling. In KRAS G12 mutants (G12C, G12D, G12), GAP binding cannot activate GTP hydrolysis; hence the signaling cascade is not terminated, and the result is an uncontrolled cell growth factor production process [Krauss]. In order to restore normal function, the present common strategy is to shift mutant KRAS molecules into the GDP-bound (inactive) conformation. However, for a long time, the RAS protein was termed "undruggable" since more than 30 years of drug development efforts were unsuccessful [8]. The "undruggability" of the RAS protein relates to the lack of binding cavities on the molecular surface, which is important in the oncogenic process. After many decades of fruitless efforts, two covalently bound KRAS G12C-mutant inhibitors were developed, called ARS-853 and ARS-1620, re-vitalizing this important research area [9-12]. Based on the success in clinical trials, Sotorasib and Adagrasib, the two irreversible inhibitors have been approved by FDA for G12C mutant lung cancer[5,13]. Very recently, the development of non-covalent inhibitors against the KRAS G12D mutant has also been addressed, and the effects of the promising MRTX-1133 compound in targeting G12D KRAS have been described [14-16] expanding the potential for the treatment of KRAS mutant tumors.

Here we present a novel strategy for finding potent new RAS inhibitors and demonstrate the potency of the method by two new molecules, inhibiting human G12D KRAS mutations. Our new molecules are not covalently bound to the G12D KRAS mutation, and, consequently, they are much more drug-like than the covalently bound ARS-853 and the ARS-1620 in the case of the G12C mutation, while their activity is comparable to them.

**Materials and Methods**

*Identification of small molecules gluing GAP and KRasG12D*

The first step of the gluing approach was to generate the coordinates of a three-dimensional molecular structure consisting of the RAS and the GAP molecules, where the two proteins were close enough for the small molecules to bind to both of them (what we call "gluing"), but sufficiently apart for allowing in the small molecules. This approach would establish novel binding sites for small, drug-like molecules *between* the "undruggable" RAS and the GAP, which the RAS structure, by itself, is lacking.

The original configuration of the RAS-GAP complex was the wild-type 1WQ1 PDB entry [17]. First, we *in silico* mutated residue 12 (glycine) in RAS (chain ID R) to aspartate in the structure 1WQ1, using the built-in software of PyMol. Next, we applied a translation by a 5 Å long vector to the GAP molecule (Figure 1). This way, an "artificial" configuration was created, where there is a suitable place for small, drug-like molecules to bind to both RAS and GAP; therefore, gluing them even in the mutated state.

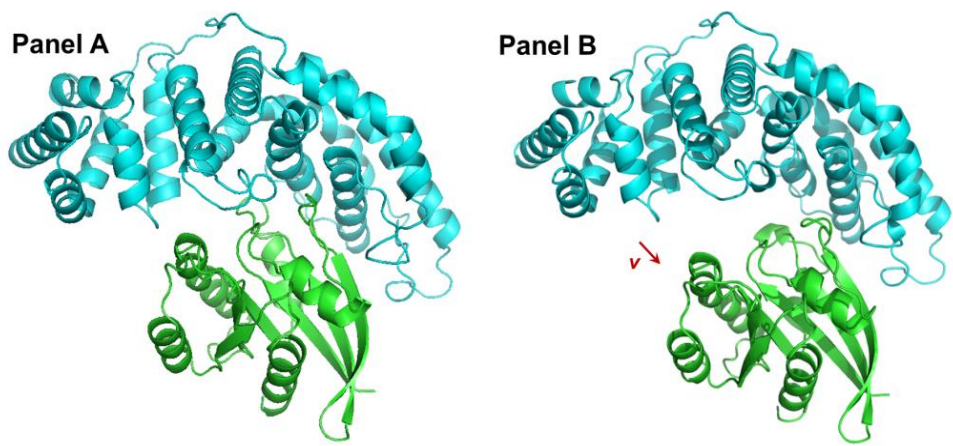

**Figure 1. Construction of the receptor complex to be used in the gluing docking.** Panel A visualizes the 1WQ1 complex structure from the PDB; proteins RAS (green) and GAP (cyan) are in the cartoon model. Panel B shows the result of the translational movement where the GAP protein molecule was translated by a 5- Å-long *v* vector (*v=(-4.903, -0.799, 0.567)*, shown as the red arrow on Panel B).

The GAP-RAS configuration of Figure 1, panel B, served as a receptor for the high-throughput *in silico* molecular docking software FRIGATE [18]. This was used for docking 4.6 million small, drug-like molecules from the "clean leads" subset of ZINC 12 database [19]. The FRIGATE docking tool minimizes the joint energy of the receptor-ligand system by combining a discrete and a continuous approach: before starting the docking procedure of millions of small molecules, it computes the potential of each possible ligand-atom in a dense rectangular grid around and in the receptor (it is the discrete step); next, by using spline-approximation from the grid points, it uses a gradient-based local optimization (it is the continuous step), enhanced by a heuristic global optimization strategy. The discrete step makes the optimization computationally feasible; the continuous step makes it efficient in finding local minima. The details of the FRIGATE tool are described in [18]. The most favorable small molecular hits are identified and listed in the molecular docking process. Fifteen of the best-scored molecules were acquired from vendors (cf Supplementary Table 1), and their anti-cancer activity was investigated in human cancer cell line cultures.

*Cell lines and culture conditions*
In experimental procedures, we used pancreatic cancer cell lines PANC-1 (KRAS$^{G12D}$) and BxPC3 (KRAS$^{wt}$), which were cultured in RPMI 1640 Medium with glutamine (Roswell Park Memorial Institute Medium, Biosera, Nuaille, France), while non-cancerous cell lines CCD-986Sk (skin fibroblast) and HUVEC-TERT (umbilical vascular endothelial) were cultured in Iscove's Modified Dulbecco's Medium (Biosera) and in Endothelial cell basal medium-2 (EBM-2; Lonza, Basel, Switzerland), respectively. RPMI and IMDM mediums were supplemented with 10% heat-inactivated Fetal Bovine Serum (FBS; Biosera) and 1% Penicillin/Streptomycin (Biosera), while CCD-986Sk cells were cultured in 20% FBS-containing medium. Medium for HUVEC-TERT cells was supplemented from the Endothelial Growth Medium-2 kit (EGM-2; Lonza) by manufacturer instructions. The cell lines were obtained from the American Type Culture Collection (ATCC). Cells were cultured in sterile T25 or T75 flasks with ventilation caps (Sarstedt, Nümbrecht, Germany) at 37 °C in a humidified atmosphere with 5% $CO_2$.

*In vitro antiproliferative activity of compounds and calculation of selectivity for KRAS$^{G12D}$ mutation*
For the evaluation of the *in vitro* antiproliferative activity of compounds, the cell viability was determined by MTT assay (3-(4,5-dimethylthiazol-2-yl)-2,5-diphenyl-tetrazolium bromide) which was obtained from Duchefa Biochemie (Haarlem, The Netherlands). After standard harvesting of the cells by trypsin-EDTA (Biosera) and phosphate-buffered saline (PBS; Biosera), $7 \times 10^3$ cells per well for pancreatic cancer cell lines and $10 \times 10^3$ cells per well for non-cancerous cell lines were seeded in 5% FBS-containing growth medium to 96-well plates with flat bottom (Sarstedt), in a 100 μL volume per well, and incubated at 37 °C. After 24 h, cells were treated with various concentrations of compounds (15 nM – 100 μM), dissolved in dimethylsulfoxide (DMSO; Sigma Aldrich, St. Louis, MO, USA, 0.5% final) and FBS-free medium (FBS final 2.5%) and incubated for 72 h under standard conditions. The control wells were treated with FBS free medium (FBS final 2.5%) and 0.5% DMSO final. Afterward, the MTT assay was performed in order to determine cell viability, by adding 22 μL of MTT solution (5 mg/mL in PBS, 0.5 mg/mL final) to each well and after 2 h of incubation at 37 °C, the supernatant was removed. The precipitated purple formazan crystals were dissolved in 100 μL of a 1:1 solution of DMSO – 96% Ethanol (Molar Chemicals Kft., Halásztelek, Hungary), and the absorbance was measured after 15 min. at λ = 570 nm by using CLARIOstar$^{plus}$ microplate reader (BMG Labtech, Ortenberg, Germany). Average background absorbance (DMSO-Ethanol) was subtracted from absorbance values of control and treated wells, and cell viability was determined relative to untreated (control) wells where cell viability was arbitrarily set to 100%. Absorbance values of treated samples were normalized versus untreated control samples and interpolated by non-linear regression analysis with GraphPad Prism 6 software (GraphPad, La Jolla, San Diego, CA, USA) to generate sigmoidal dose-response curves from which the 50% inhibitory concentration ($IC_{50}$) values of the compounds were calculated, and presented as micromolar (µM) units. The experiments were done in triplicate, and each experiment was repeated twice. Selectivity of the compounds toward $KRAS^{G12D}$ mutation compared to $KRAS^{wt}$ is calculated based on the ratio between $IC_{50}$ of $KRAS^{wt}$ or non-cancerous cell line and $IC_{50}$ of $KRAS^{G12D}$ mutated cell line. Selectivity values higher than 1 represent selectivity toward $KRAS^{G12D}$ mutation compared to $KRAS^{wt}$, while values lower than 1 represent the opposite.

*Experimental animals*
Adult female BALB/c mice were used in the chronic toxicity study, while the immunodeficient SCID male mice were used in subcutaneous PANC-1 and BxPC3 human pancreatic tumor models *in vivo* experiments. Mice were held in filter-top boxes in the experimental barrier rooms, and every box opening was performed under a laminar-flow hood in sterile conditions. The cage components, corn-cob bedding, and food (VRF1 from Special Diet Services) were steam-sterilized in an autoclave (121 °C, 20 min). The distilled water was acidified to pH 3 with hydrochloric acid. The animals used in these studies were cared for according to the "Guiding Principles for the Care and Use of Animals" based upon the Helsinki Declaration, and they were approved by the local ethical committee. The animal housing density was according to regulations and recommendations from directive 2010/63/EU of the European Parliament and of the Council of the European Union on the protection of animals used for scientific purposes. Permission license for breeding and performing experiments with laboratory animals: PEI/001/1738-3/2015 and PEI/001/2574–6/2015.

*Chronic toxicity study of compound 14*
In order to determine the toxicity of compound **14** on healthy animals, a chronic toxicity study was performed. Adult BALB/c female mice (18–20 g), which were kept under the conditions as described above, were treated with compound **14** by intraperitoneal (i.p.) administration with a dose of 6 mg/kg in a volume 0.1 ml per 10 g of mice body weight, on days 1, 3, 5, 8, 10 and 12. In the case of the control group, 8% Ethanol (Molar Chemicals Kft.) / 8% DMSO (Sigma Aldrich) in sterile water for injection (Pharmamagist Kft., Budapest, Hungary) as solvent was administered. Each group consisted of three mice. The toxicity was evaluated on the basis of life span, liver toxicity, behavior and appearance of the mice, as well as the body weight. Parameters were followed for 15 days.

*Mouse models of subcutaneous human pancreatic cancers PANC-1 ($KRAS^{G12D}$) and BxPC3 ($KRAS^{wt}$), doses and schedule of compound 14 treatments, and measurements*

Adult SCID male mice (32–41 g) were used in this experiment and kept under the conditions described above. PANC-1 (KRAS$^{G12D}$) and BxPC3 (KRAS$^{G12D}$) human pancreatic cancer cells were injected subcutaneously into mice, whereby 3 x 10$^6$ cells were used per animal, suspended in 200 µL of RPMI medium (Biosera). Treatments started 23 and 7 days after cells inoculation, when average tumor volume was 64.0 and 55.4 mm$^3$, respectively, for PANC-1 and BxPC3 tumor-bearing mice. Compounds were dissolved in 8% Ethanol (Molar Chemicals Kft.) and 8% DMSO (Sigma Aldrich) in sterile water for injection (Pharmamagist Kft.) solution and administered by i.p. injection in a volume 0.1 ml per 10 g of mice body weight, 3 times per week. For the PANC-1 model 2 groups with 9 animals per group, while in BxPC3 model 2 groups with 10 animals per group were established. The mice in the control group were treated with the solvent. Animals bearing PANC-1 tumor were treated by the next schedule and doses: with 6 mg/kg in 8% Ethanol/8% DMSO/water on days 24, 27, 29, 31, 34, 36, 38, 41, 43, 45, 48, 50, 55, 57, 59, 62, 64, 66, 69, 71, 73, 76 and 78 after cell inoculation. Animals bearing BxPC3 tumor were treated by the next schedule and doses: with 6 mg/kg in 8% Ethanol/8% DMSO/water on days 8, 11, 13, 15, 18, 21, 25, 28, 32, 35, 39, 41, 43, 46 and 48 after cell inoculation. Tumor volumes were measured initially when the treatment started and at periodic intervals. A digital caliper was used to measure the longest (a) and the shortest diameter (b) of a given tumor. The tumor volume was calculated using the formula V = ab$^2$ × π/6, whereby a and b represent the measured parameters (length and width). The termination of the experiment was 80 days after cell inoculation, i.e., 57 days after treatment started for the PANC-1 model, and 53 days after cell inoculation, i.e., 46 days after treatment started for the BxPC3 model, since the average volume of the tumors in the control (PANC-1) and in compound **14** (BxPC3) groups reached over 1600 mm$^3$. The mice from all groups were sacrificed by cervical dislocation, after which their tumors and livers were harvested.

The antitumor effect of compound **14** was evaluated by measuring the tumor volume and calculating the percentage of how much tumor volume grew in comparison to the starting tumor volume, which was set for all tumors as 100% at the start of treatment. Additionally, the growth-rate coefficient of the tumor growth was determined using non-linear fitting for each tumor and average was calculated for each group. The doubling time of the tumor was calculated similarly. The toxicity effects of the compound were evaluated by measuring the animal body and liver weights and calculating the liver weight/body weight ratio.

## Results and Discussion

### *Identification and characteristics of the candidate molecules from the gluing docking strategy*

The gluing strategy, with the artificially created gapped KRAS-GAP molecular structure, to which we have performed a high-throughput in silico molecular docking experiment, is demonstrated on the panels of Figure 1. Supplementary Table 1 provides characteristics of the 15 candidate compounds. The structures and the docking sites of the 15 candidate compounds are shown in Figure 2.

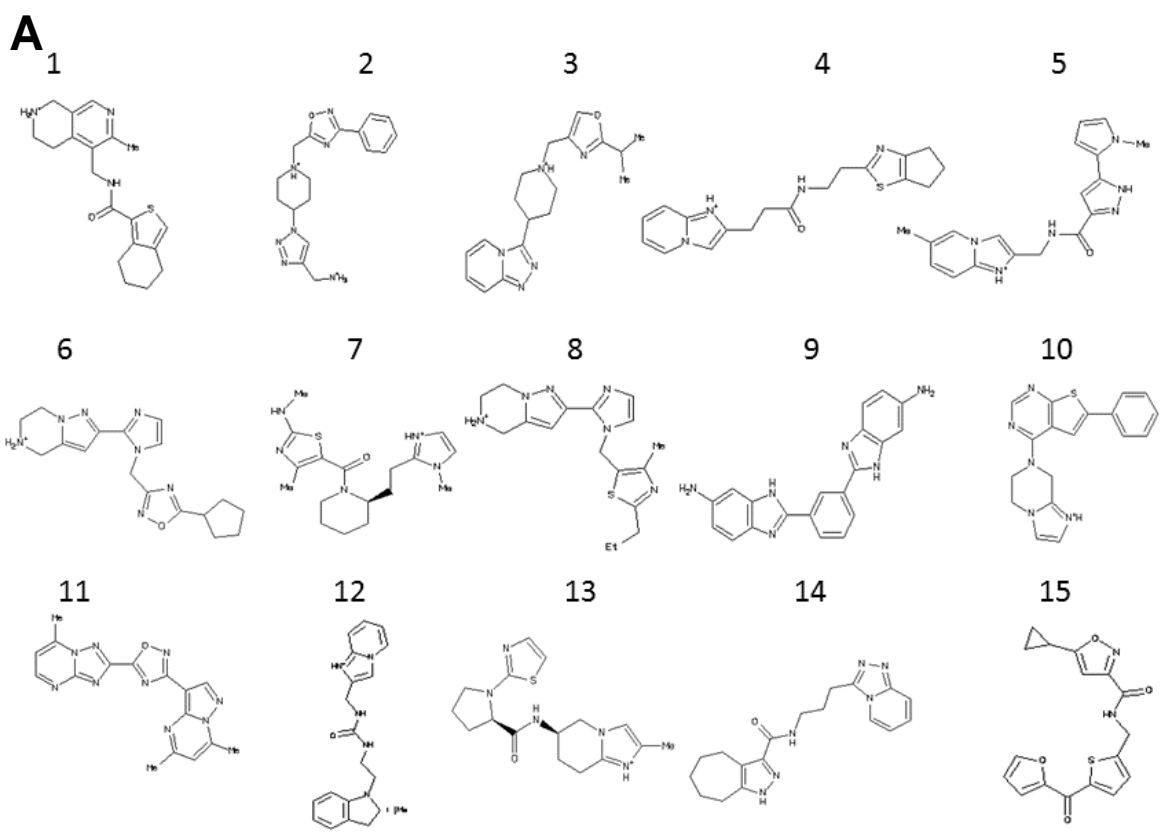

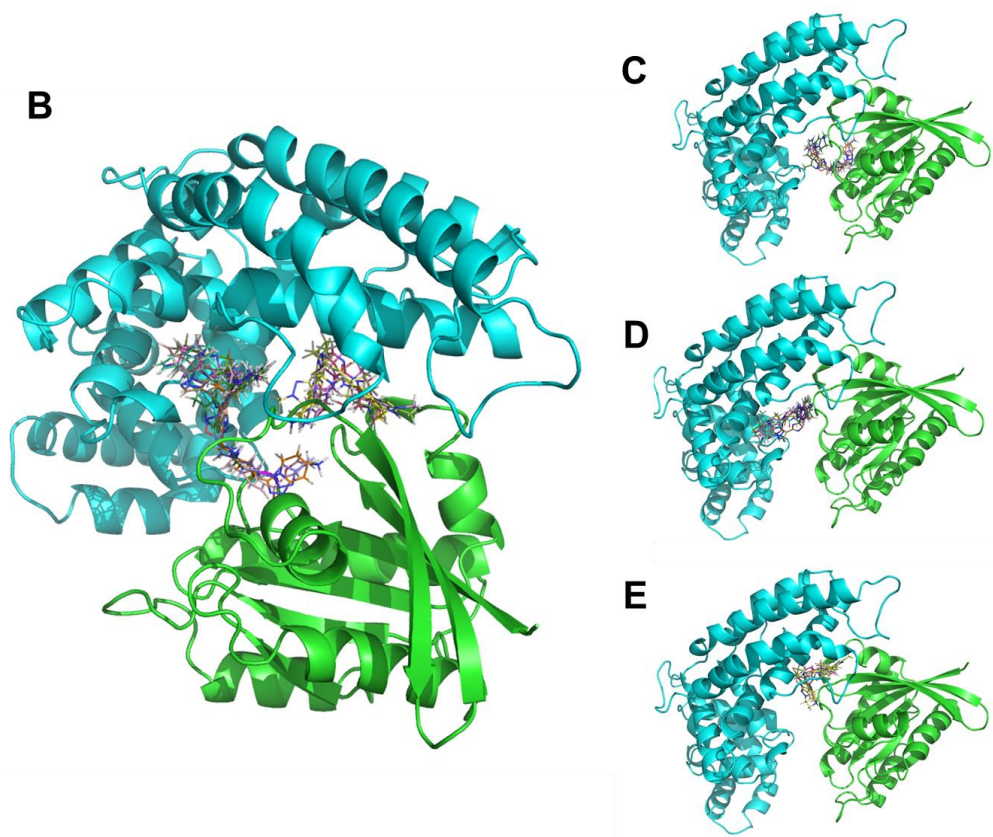

**Figure 2. Structural formulae (Panel A) and docking conformations (Panels B, C, D, E) of the 15 candidate compounds, as identified during the gluing docking method.** Proteins RAS (green) and GAP (cyan) are shown in the cartoon model, while docked compounds are shown in lines. The figure was created by Pymol (the Pymol file is provided in the Supplementary Data Set 1).

Figure 2 shows that there are three rather separate docking pockets for the 15 compounds that involve different sites in the RAS and GAP proteins. The three separate docking pockets are shown one by one on Figure 2 panels C, D and E. Compounds **2, 3, 8,** and **15** are grouped together on panel C, while compounds **4, 5, 9, 11, 12,** and **14** are shown together on panel D. Finally, Panel E represents a third docking site populated with compounds **1, 6, 7, 10,** and **13**.

### *In vitro antiproliferative activity of the compounds and selectivity toward KRAS$^{G12D}$ mutation*

**Table 1** shows the results of *in vitro* screening of the 15 compounds on PANC-1 (KRAS$^{G12D}$) and BxPC3 (KRAS$^{wt}$) human pancreatic cancer cell lines. In these experiments, compounds **10** and **14** showed the highest antitumor activity with IC$_{50}$ values of 2.3 μM and 5.5 μM, respectively, on the PANC-1 cell line. Importantly, these two compounds also showed considerable specificity ratios, with a decreased IC$_{50}$ value on the KRAS$^{G12D}$-expressing cell line, as compared to the KRAS$^{wt}$ - expressing cell line (**Table 1**). We have also observed that at lower drug concentrations, several additional compounds also showed higher antiproliferative activity against the KRAS$^{G12D}$ cell line PANC-1 (see the IC$_{50}$ values in Table 1). Based on the IC$_{25}$ values, the specificity levels of compounds **10** and **14** were also considerably increased (**Table 1**).

It was of immediate interest to also decide if the newly identified compounds show similar or even better effects in the in vitro tests as compared to the recently described KRAS$^{G12D}$ allele-specific inhibitor, **MRTX-1133** [14]. We have observed that compounds **10** and **14** showed 8.1-fold and 3.4-fold higher antitumor activity, respectively, as compared to **MRTX-1133**. In addition, allele-specific selectivity for PANC-1 KRAS$^{G12D}$ mutation was lower for MRTX-1133 as compared to the new compounds **10** and **14** (**Table 1**).

**Table 1**. **Antiproliferative effect of 15 compounds and MRTX-1133 on PANC-1 (KRAS$^{G12D}$) and BxPC3 (KRAS$^{wt}$) human pancreatic cancer cell lines and their selectivity toward KRAS$^{G12D}$ mutation.** IC$_{50}$ and IC$_{25}$ values in average ± SD are only for compounds **10**, **14,** and **MRTX-1133** because the other compounds showed very low potency and low selectivity toward KRAS$^{G12D}$ mutation in the first screening, hence these were not further tested. Selectivity represents the ratio between IC$_{50}$BxPC3$^{KRAS-wt}$ and IC$_{50}$PANC-1$^{KRAS-G12D}$ values. Selectivity values higher than 1 represent selectivity toward KRAS$^{G12D}$ mutation compared to KRAS$^{wt}$, while values lower than 1 represent the opposite.

| Compound name | IC$_{50}$ (µM) | | Selectivity toward KRAS$^{G12D}$ mutation | IC$_{25}$ (µM) | | Selectivity toward KRAS$^{G12D}$ mutation |
| --- | --- | --- | --- | --- | --- | --- |
| | PANC-1 (KRAS$^{G12D}$) | BxPC3 (KRAS$^{wt}$) | | PANC-1 (KRAS$^{G12D}$) | BxPC3 (KRAS$^{wt}$) | |
| **1** | 53.77 | 48.45 | 0.9 | 17.36 | 30.99 | 1.8 |
| **2** | 69.13 | 35.05 | 0.5 | 16.09 | 19.35 | 1.2 |
| **3** | 95.09 | 42.30 | 0.4 | 31.14 | 24.84 | 0.8 |
| **4** | 93.69 | 39.91 | 0.4 | 28.40 | 22.67 | 0.8 |
| **5** | 118.10 | 30.76 | 0.3 | 32.84 | 14.85 | 0.5 |
| **6** | 93.52 | 45.08 | 0.5 | 41.55 | 27.30 | 0.7 |
| **7** | 138.60 | 40.55 | 0.3 | 52.20 | 24.16 | 0.5 |
| **8** | 90.71 | 43.73 | 0.5 | 26.84 | 27.07 | 1.0 |
| **9** | 74.12 | 23.23 | 0.3 | 31.14 | 13.44 | 0.4 |
| **10** | 2.3 ± 0.5 | 3.8 ± 0.1 | 1.7 | 0.04 ± 0.02 | 0.6 ± 0.001 | 19.3 |
| **11** | 117.30 | 43.18 | 0.4 | 50.28 | 25.30 | 0.5 |
| **12** | 93.11 | 42.97 | 0.5 | 31.41 | 25.12 | 0.8 |
| **13** | 103.20 | 47.54 | 0.5 | 41.12 | 29.96 | 0.7 |
| **14** | 5.5 ± 1.8 | 7.9 ± 0.1 | 1.5 | 1.2 ± 0.5 | 4.1 ± 0.4 | 3.5 |
| **15** | 78.40 | 55.75 | 0.7 | 20.72 | 35.15 | 1.7 |
| **MRTX-1133** | 18.6 ± 3.3 | 20.1 ± 5.4 | 1.1 | 6.0 ± 3.3 | 12.4 ± 3.8 | 2.1 |

Furthermore, the two most potent and selective compounds, **10** and **14**, were also investigated on two different non-cancerous cell lines in order to determine their selectivity in comparison to normal cell lines (**Table 2**). These data showed, with the exception of compound **10** on HUVEC-TERT cells, that both compounds are less potent against the non-cancerous cell lines HUVEC-TERT (umbilical vascular endothelial) and CCD-986Sk (skin fibroblast) with higher $IC_{50}$ values compared to PANC-1, and with higher selectivity for $KRAS^{G12D}$ mutation. Compound **14** showed lower potency against non-cancerous cell lines, especially on skin fibroblast cells, suggesting that it may not be toxic for normal cells and it is selective for cancer cell lines.

**Table 2. Antiproliferative effect of compounds 10 and 14 on non-cancerous cell lines HUVEC-TERT (umbilical vascular endothelial) and CCD-986Sk (skin fibroblast) and their selectivity toward $KRAS^{G12D}$ mutation in comparison to non-cancerous cells.** $IC_{50}$ values are shown as (average ± SD). Selectivity represents the ratio of the $IC_{50}$ values determined for non-cancerous cell lines and the PANC-$1^{KRAS-G12D}$ cell lines. Selectivity values higher than 1 represent selectivity toward $KRAS^{G12D}$ mutation compared to non-cancerous cells, while values lower than 1 represent the opposite.

| Non-cancerous cell line | compound 10 | | compound 14 | |
|---|---|---|---|---|
| | $IC_{50}$ (µM) | Selectivity toward $KRAS^{G12D}$ mutation | $IC_{50}$ (µM) | Selectivity toward $KRAS^{G12D}$ mutation |
| HUVEC-TERT | 0.3 ± 0.02 | 0.1 | 8.9 ± 0.6 | 1.6 |
| CCD-986Sk | 6.8 ± 1.6 | 3.0 | 26.1 ± 4.1 | 4.7 |

*Compound 14 shows selective in vivo inhibition of tumor growth against $KRAS^{G12D}$ xenograft - expressing mice*

Encouraged by the high antitumor potency of compounds **10** and **14** and their good selectivity toward $KRAS^{G12D}$ cancer cell line in comparison to the $KRAS^{wt}$ cancer and non-cancerous cell lines, we have initiated *in vivo* efficacy studies. However, during the estimation of the appropriate doses, compound **10** could not be dissolved for administration into the animals in suitable dose, while compound **14** showed acceptable solubility for doses up to 6 mg/kg. Based on the solubility parameters, and the additional earlier observation that compound **14** showed better selectivity toward cancer than to non-cancerous cell lines as compared to compound **10**, compound **14** was chosen for the *in vivo* investigation.

**Chronic toxicity study of compound 14 *in vivo***
In order to decide whether the dose and treatment schedule for *in vivo* studies are toxic or not, *in vivo* chronic toxicity study with 6 injections of the treatment in a dose of 6 mg/kg was performed on healthy animals to determine the toxicity of compound **14**. After 15 days, the general appearance and behavior of experimental animals were adequate, while no significant change in body weight (body weight increased in control and treated group for 4.0 and 7.4% respectively) (**Figure 3A**), as well as liver weight/body weight ratio could be observed (**Figure 3B**), suggesting that compound **14** is a non-toxic substance that it can be further investigated on tumor-bearing mice.

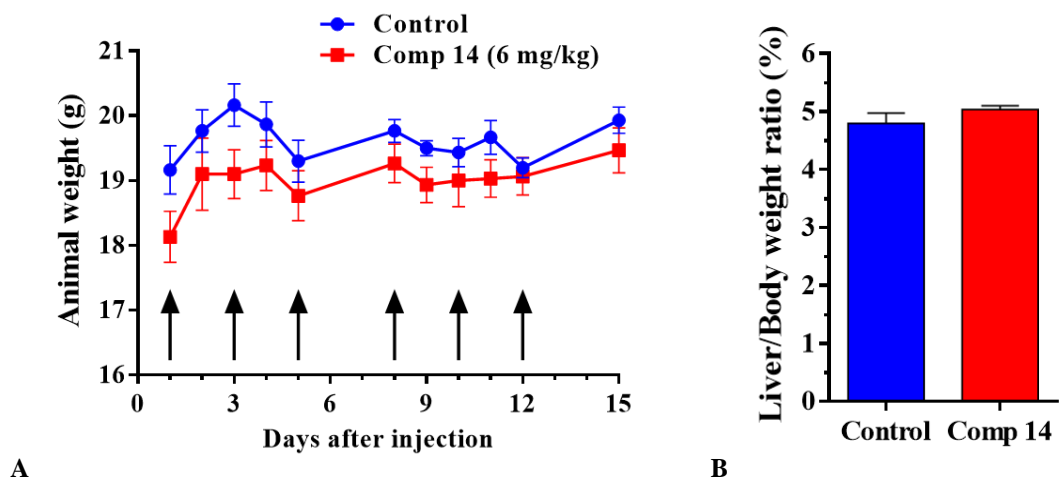

**Figure 3.** *In vivo* **chronic toxicity study of compound 14 on healthy BALB/c female mice with 6 treatments (black arrows) under a dose of 6 mg/kg**. (**A**) Animal body weight (grams, average ± SEM). (**B**) Liver weight/body weight ratio (percentage, average ± SEM) after the termination of the experiment, 15 days subsequent to the first treatment. 3 animals were used per group. Statistical analysis was performed by Mann–Whitney test.

## Effect of compound 14 in subcutaneous human pancreatic PANC-1 (KRAS$^{G12D}$) and BxPC3 (KRAS$^{wt}$) tumor models *in vivo*

We have shown that compound **14** has an antitumor activity and selectivity toward KRAS$^{G12D}$ cancer cell line PANC-1 in comparison to the KRAS$^{wt}$ cancer and non-cancerous cell lines. It is also shown that the solubility of compound **14** is sufficient for further studies, and it is not toxic for animals. With these results, we progressed towards investigating the *in vivo* antitumor activity and specificity of compound **14** on xenograft mouse models using immunodeficient SCID male mice bearing KRAS$^{G12D}$ (PANC-1) or KRAS$^{wt}$ (BxPC3) tumors.

Based on animal body weight, which was not significantly changed in both groups and in both models, during the experiment, it was shown that compound **14** is not toxic for experimental animals during the treatment (**Figure 4A, Figure 5A, Table 3**). Moreover, the decrease of the body weight was higher in the control group as compared to compound **14** treated group in both models. In addition to body weight, non-significant changes in liver weight/body weight ratios confirmed non-toxicity of this compound for experimental animals in both models (**Figure 4B, Figure 5B, Table 3**).

Regarding antitumor activity, results indicated that compound **14** reduced tumor volume represented in mm$^3$ as compared to the control group in KRAS$^{G12D}$ mutated model of PANC-1 by 13.1% (**Figure 4C, Table 3**), while in KRAS$^{wt}$ model of BxPC3 tumor volume was increased for 11.1% in the compound **14** treated group compared to the control (**Figure 5C, Table 3**), suggesting the selectivity of compound **14** *in vivo* toward the KRAS$^{G12D}$ mutation.

This antitumor activity and selectivity were confirmed when evaluating the tumor volume represented in percentage. Setting all tumor volumes as 100% at the start of treatment, and following their growth in percentage, it was observed that compound **14** inhibited tumor volume growth as compared to tumor volume in the control group for 23.8% in PANC-1 model (**Figure 4D, Table 3**), while increased tumor volume for 22.8% in BxPC3 model (**Figure 5D, Table 3**).

Additionally, determining the growth-rate coefficient of the tumor growth using non-linear fitting for each tumor, it was observed that compound **14** inhibited tumor growth rate by 26.3% in comparison to the control group (p=0.1348) in KRAS$^{G12D}$ mutated model of PANC-1 (**Figure 4E, Table 3**), while the growth-rate coefficient increased by the treatment in KRAS$^{wt}$ model of BxPC3 for 9.9% (p=0.0887) compared to the control group (**Figure 5E, Table 3**). Moreover, the antitumor activity and selectivity toward KRAS$^{G12D}$ mutation *in vivo* were confirmed by calculating the time required for tumor doubling in size. This parameter increased during treatment with compound **14** by 14.5% (p=0.1348) in KRAS$^{G12D}$ mutation PANC-1 bearing tumor model (**Figure 4F, Table 3**), while decreased by 12.1% (p=0.0887) in the tumor-bearing KRAS$^{wt}$ model (**Figure 5F, Table 3**), revealing that compound **14** slowed down the progression of KRAS$^{G12D}$ mutated tumor as compared to KRAS$^{wt}$ tumor *in vivo*.

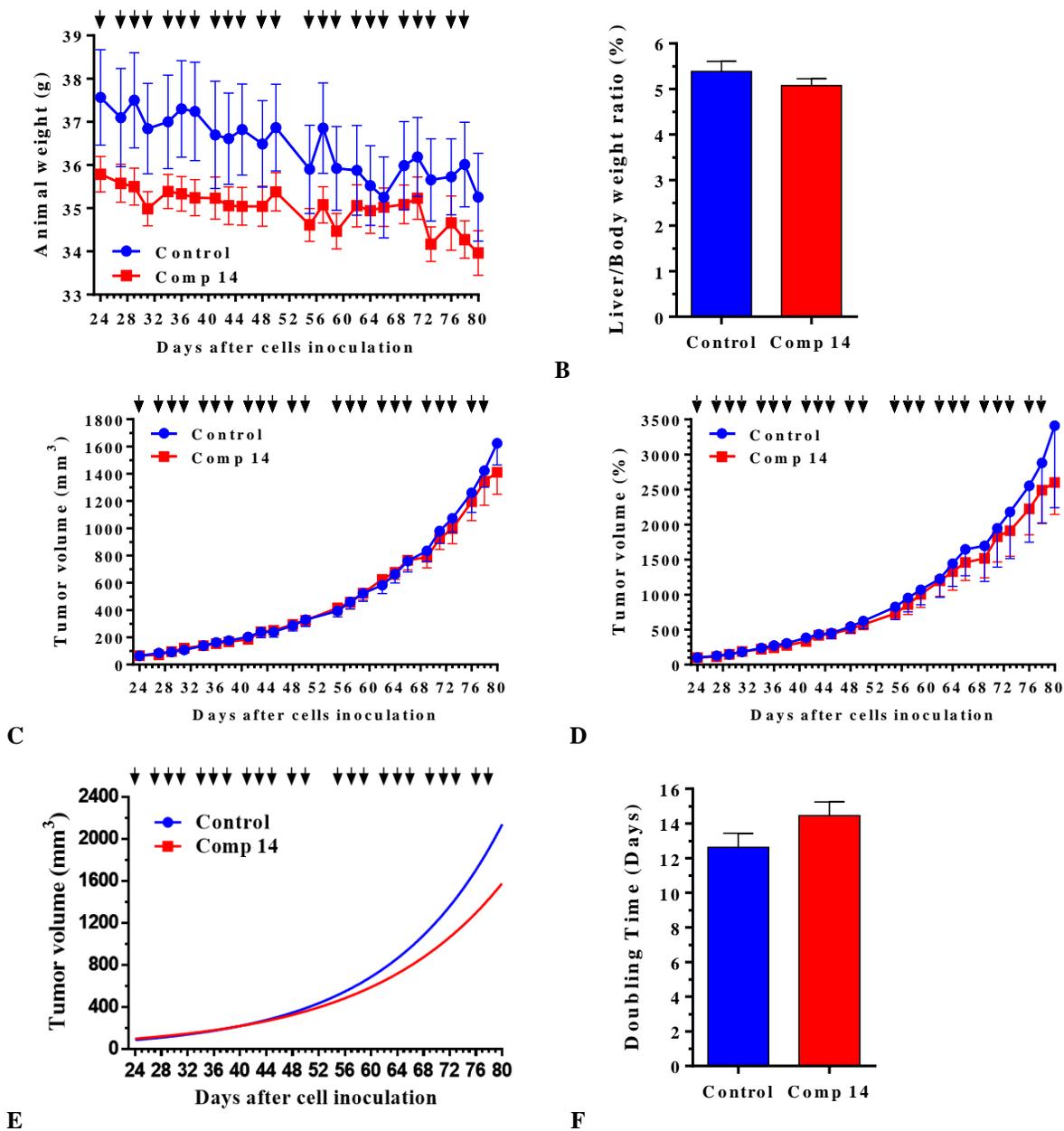

**Figure 4. Effect of compound 14 (6 mg/kg, black arrows) in subcutaneous human pancreatic PANC-1 (KRAS$^{G12D}$) tumors bearing SCID male mice.** (A) Animal body weight (grams, average ± SEM). (B) Liver weight/body weight ratio (percentage, average ± SEM) after termination of the experiment, 80 days after cell inoculation. (C) Tumor volume (mm$^3$, average ± SEM). (D) Tumor volume (percentage, average ± SEM). (E) Tumor volume by non-linear fitting (mm$^3$, average). (F) Tumor doubling time (days, average ± SEM). 9 animals were used per group.

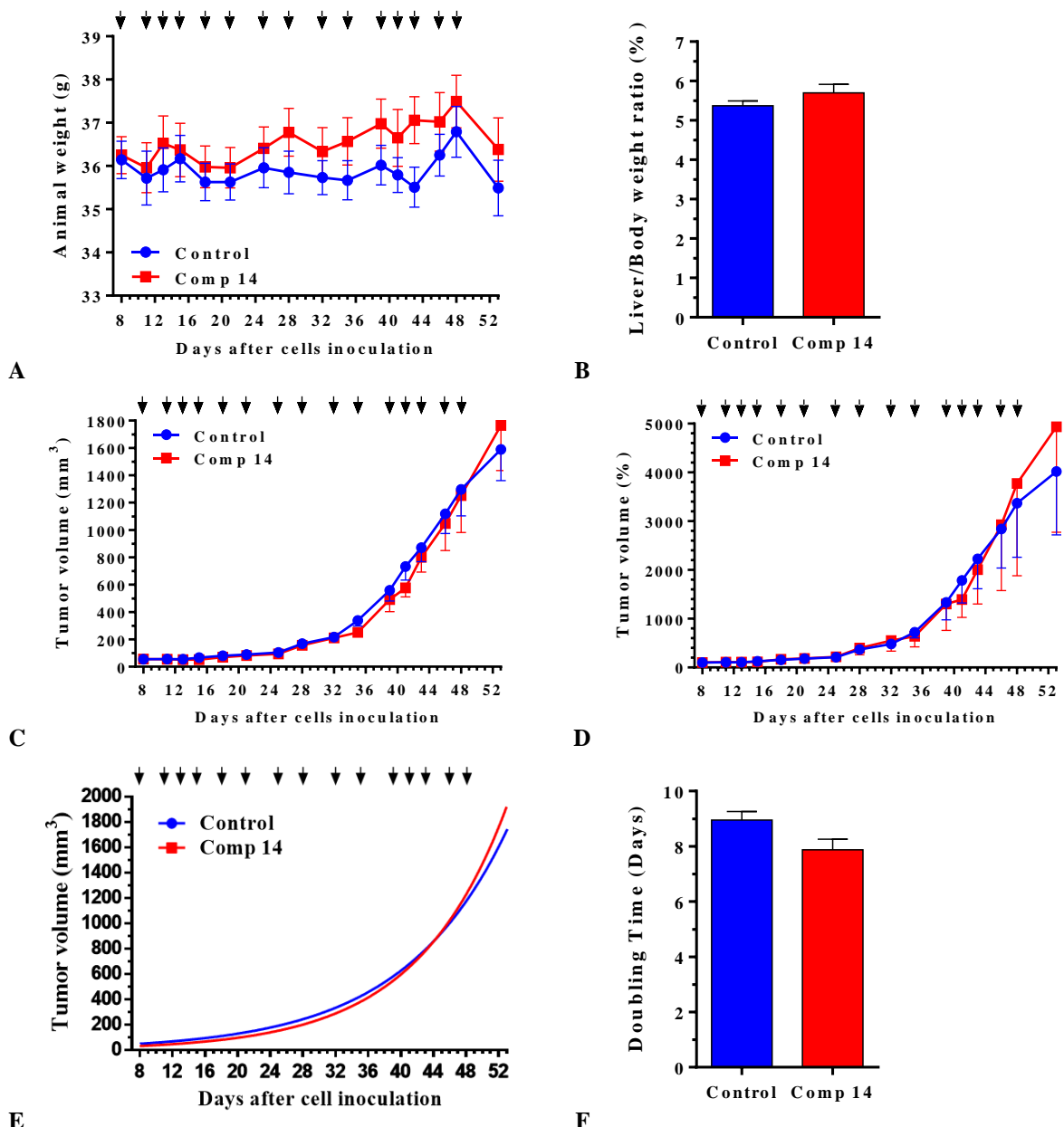

**Figure 5**. Effect of compound **14** (6 mg/kg, black arrows) in subcutaneous human pancreatic BxPC3 (KRAS$^{wt}$) tumors bearing SCID male mice. (A) Animal body weight (grams, average ± SEM). (B) Liver weight/body weight ratio (percentage, average ± SEM) after termination of the experiment, 53 days after cell inoculation. (C) Tumor volume (mm$^3$, average ± SEM). (D) Tumor volume (percentage, average ± SEM). (E) Tumor volume by non-linear fitting (mm$^3$, average). (F) Tumor doubling time (days, average ± SEM). 10 animals were used per group.

**Table 3. Effect of compound 14 in subcutaneous human pancreatic PANC-1 (KRAS$^{G12D}$) and BxPC3 (KRAS$^{wt}$) tumors bearing mice.** Data represent percentage values (%) where negative values indicate a decrease of the animal body weight at the end of the experiment, compared to the start; a decrease of the liver weight/body weight ratio compared to the control group; inhibition of tumor-growth, compared to the control group; and decrease of the time for tumor-doubling, compared to the control group.

| Parameter | Control | | Comp 14 | |
|---|---|---|---|---|
| | PANC-1 | BxPC3 | PANC-1 | BxPC3 |
| Animal body weight | -6.1 | -1.8 | -5.1 | +0.4 |
| Liver/Body weight ratio | | | -5.8 | +5.8 |
| Tumor volume in mm$^3$ | | | -13.1 | +11.1 |
| Tumor volume in % | | | -23.8 | +22.8 |
| Tumor volume by non-linear fitting | | | -26.3 | +9.9 |
| Tumor doubling time | | | +14.5 | -12.1 |

The inhibition of tumor growth by compound **14** seems to be selective against KRAS$^{G12D}$ expressing tumor model, as in KRAS$^{wt}$ tumor-bearing mice inhibition of the tumor growth by the same treatment was not observed. On the contrary, treatment by compound 14 had deleterious effects on the KRAS$^{wt}$ tumor-bearing mice. Further support for this statement is provided by comparing the growth-rate coefficient and doubling time, as independent parameters, of the compound **14** treatments in both models. In these data, we obtain a significant difference (p < 0.0001).

*Molecular interactions of compounds 10 and 14 with the complex of GAP and G12D mutant KRAS*

In order to investigate the detailed molecular interactions of the most efficient compounds with the complex of GAP and G12D mutant KRAS, Figures 6 and 7 provide a close-up of the docking sites.

In our structural model, compound **10**, consisting of an imidazo-pyrazinyl a thyenopyrimidie and a phenyl group, is situated between the interaction surfaces of RAS and GAP (Figure 6). The molecule fits into the hydrophobic pocket between the C-terminal end of Switch-II and the loop between β-5 strand and α-3 helix of RAS and into the hydrophobic gap between the α21 and α22 helices of GAP.
The pyrazole and imidazole rings of the molecule are situated in the hydrophylic pocket of KRAS formed by the main and side chains of residues E37, G60, Q61, E63, S65, and R68 that create a negatively charged protein surface patch. The NH$^+$ group of the imidazole group is in an H-bond with the main chain carboxyl group of E63, while the pyrimidine ring is placed in the negatively charged pocket between the α21 and α22 helices of the GAP protein. The phenyl group of the compound faces the apolar residues of the loop between the α14 and α15 helices of GAP (residues F901, L902, L904, I905), as well as the main chain of the 60-63 residues of KRAS.

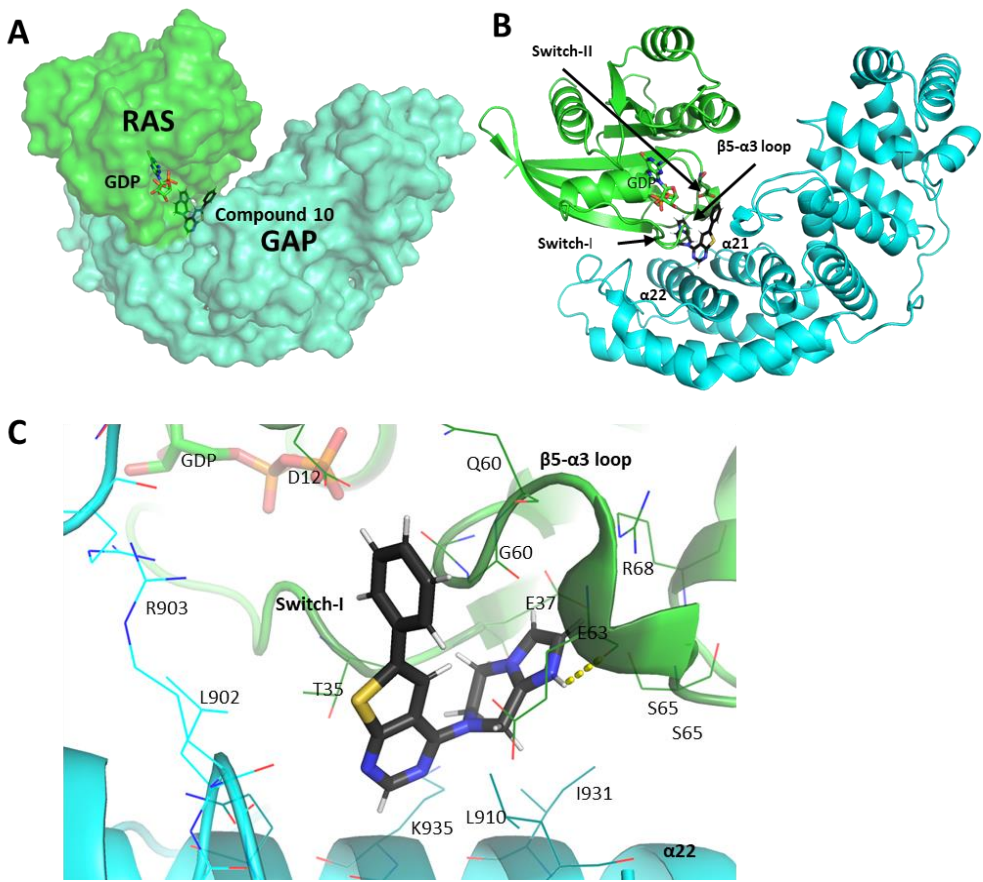

**Figure 6. Docking site of compound 10 in the structural model of KRAS-G12D and GAP. A)** shows compound **14** upon the protein surface, while **B)** shows the compound with the secondary/tertiary structural elements of the proteins. **C)** is a close-up image of the binding site, with the nearby, potentially interacting residues shown as sticks. RAS is shown in green; GAP is shown in cyan. Darker shades on the cartoons and sticks indicate residues within 4 Å of the compound. Legends in bold indicate secondary structural elements, or domains of the proteins, while residues are indicated by light legends. H-bonds are shown as yellow dashed lines. Figures were created by PyMol visualization tool. The RAS-GAP configuration and the molecule docking process were done as described in the Methods section.

Compound **14** is located in the KRAS-GAP interaction site nearby the nucleotide-binding pocket (Figure 7). The triazole and the propyl moieties of the compound are aligned between Switch-I and the turn motif between the α22 helix and β20 strand of GAP, where the triazole ring interacts with the hydrophobic side chains of Switch-I and the carbon atoms of the propyl moiety are surrounded by some apolar residues from GAP. The pyrazole and cyclohexane groups of the compound mainly interact with the GAP surface. The pyrazole ring is placed in between the guanidine groups of R789 and R894 while forming an H-bond with T785-residue of the GAP protein, and the cycloheptane ring is placed in the pocket formed by the hydrophobic sidechains of residues at the α13 and α19 helices of GAP (L787, F788, M891, V895, L902).

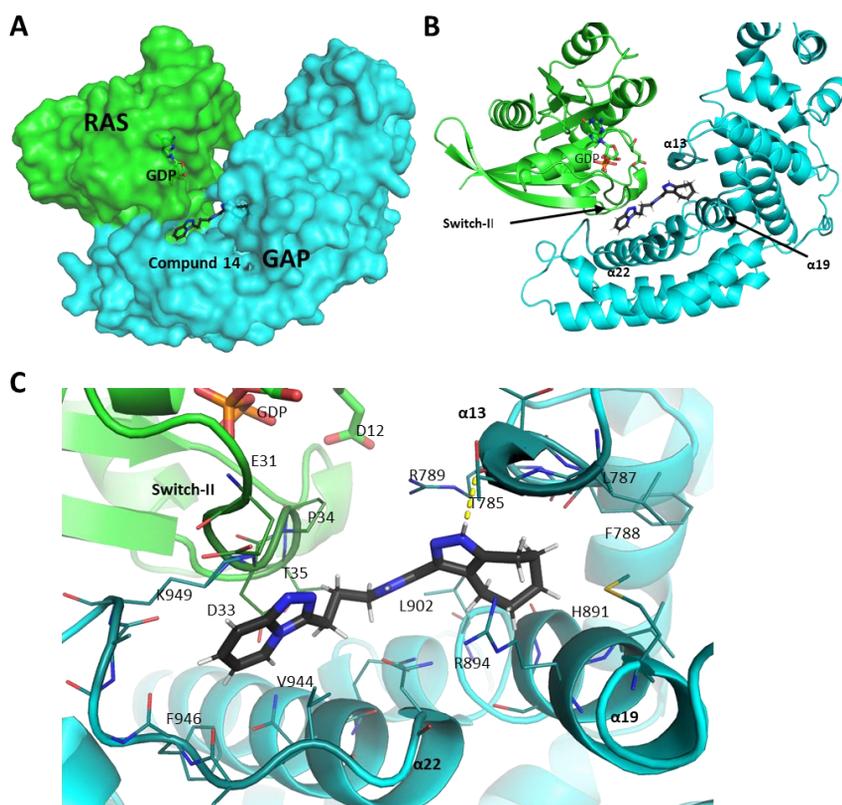

**Figure 7. Docking site of compound 14 in the structural model of KRAS-G12D and GAP. A)** shows Molecule-14 in relation to the protein surface, while **B)** shows the compound in relation to the secondary structural elements of the proteins. **C)** is a close-up image of the binding site, with the nearby, potentially interacting residues shown as sticks. KRAS is shown as green; GAP is shown as blue. Darker shades on the cartoons and sticks indicate residues within 4 Å of the compound. Legends in bold indicate secondary structural elements, or domains of the proteins, while residues are indicated by light legends. Figures were created by PyMol visualization tool. The KRAS-GAP configuration and the molecule docking process were done as described in the Methods section.

**Conclusions**

In order to inhibit uncontrolled cell growth, we established a new method for finding small molecules, binding to both the GAP and the mutated KRAS$^{G12D}$ molecules, gluing them together, thus serving as novel drug candidates for innovative cancer therapies. By this novel method in *in vitro* screening, we identified and selected two small molecular drugs, compounds **10** and **14**, which specifically and selectively inhibit the growth of human pancreatic cancer cell line with KRAS$^{G12D}$ mutation compared to KRAS$^{wt}$ cancer and normal cell lines, with higher capacity then **MRTX-1133**.

Moreover, the inhibition of tumor growth *in vivo* by compound **14** under a dose of 6 mg/kg, 2-3x/week seems to be selective against PANC-1 KRAS$^{G12D}$ mutated pancreatic tumor model compared to BxPC3 KRAS$^{wt}$ tumor-bearing mice, with no toxicity and side effects for the experimental animals in either model. For higher tumor inhibition and selectivity, further studies will need to increase the solubility of compound **14** to allow administration of doses higher than 6 mg/kg. Alternatively, an increased frequency of administration and/or different administration routes are to be considered.

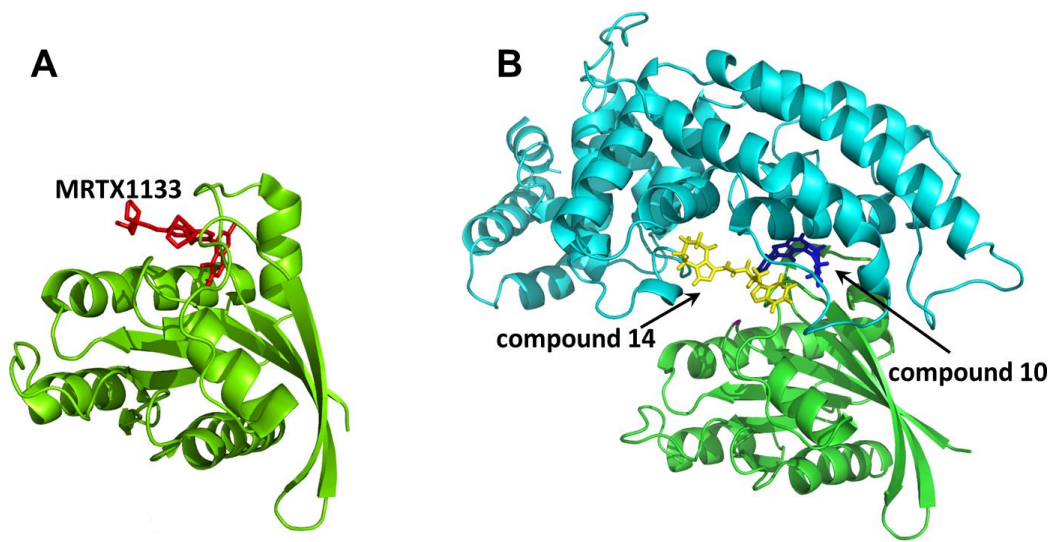

**Figure 8. Binding of MRTX-1133 as compared to the docking sites of compounds 10 and 14.** Panel A shows the RAS$^{G12D}$:MRTX1133 complex (PDB ID: 7RPZ [14]), the RAS protein is in green cartoon model, the MRTX1133 molecule is in red sticks. Panel B represents the docking positions of the newly identified compounds **10** (blue sticks) and **14** (yellow sticks) within the receptor RAS$^{G12D}$:GAP complex (created as described in the Methods). GAP is in cyan cartoon model; RAS is in green cartoon model.

Our present results indicate that the large-scale and high-throughput docking may be useful when paired with the novel gluing strategy to identify promising compounds against mutant KRAS. It is of great importance to note that the binding mode of the previously suggested G12D-specific

MRTX-1133 compound is associated with different characteristics as compared to the binding observed in the structural model of our new compounds **10** and **14 (Figure 8)**. This observation suggests that further studies for inhibitor design may also focus on an expanded surface of the GAP:RAS complex.

**Funding**

Supported by the National Research, Development and Innovation Fund of Hungary (PD134324 to KN and K109486, K135231, FK137867, NKP-2018-1.2.1-NKP-2018-00005, 2022-1.2.2-TÉT-IPARI-UZ-2022-00003 to BGV), and the TKP2021-EGA-02 grant, implemented with the support provided by the Ministry for Innovation and Technology of Hungary from the National Research, Development and Innovation Fund. KN was also supported by the Parents Back to Science program of the Budapest University of Technology and Economics (BME) and the ÚNKP-20-4 (ÚNKP-20-4-II-BME-311) New National Excellence Program of the Ministry for Innovation and Technology from the source of the National Research, Development and Innovation Fund. VG was partially funded by the Ministry of Innovation and Technology of Hungary from the National Research, Development and Innovation Fund, under the ELTE TKP 2021-NKTA-62 funding scheme.

## Supplementary Table

Table S1: The fifteen experimentally examined molecules, with ZINC numbers, suppliers, catalogue numbers, structures and molecular weights are listed here. The molecules are referred by their numbers in the first column in the main text. The ZINC numbers refer to the https://zinc.docking.org resource.

| ID in main text | ZINC ID | Catalogue number | Structure | MW (g/mol) |
|---|---|---|---|---|
| 1 | 72410037 | ChemBridge 29004287 | 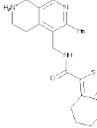 | 342 |
| 2 | 23339127 | ChemBridge 43533033 | 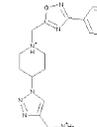 | 341 |
| 3 | 97632085 | ChemBridge 59994951 | 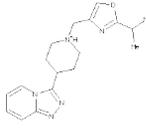 | 326 |
| 4 | 65383408 | ChemBridge 27105214 | 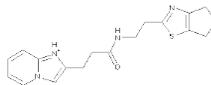 | 341 |
| 5 | 91526788 | ChemBridge 37041256 | 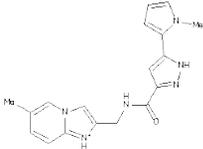 | 335 |
| 6 | 72419308 | ChemBridge 16262498 | 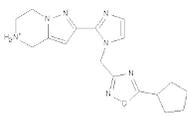 | 340 |
| 7 | 77385938 | ChemBridge 42971154 | 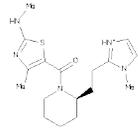 | 348 |

| 8 | 72407267 | ChemBridge 11188583 | 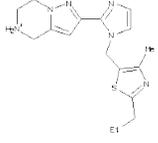 | 343 |
|---|---|---|---|---|
| 9 | 348992 | ChemDiv 0263-0451 | 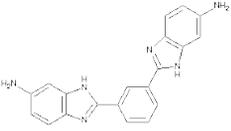 | 340 |
| 10 | 97201518 | ENAMINE Z991000544 | 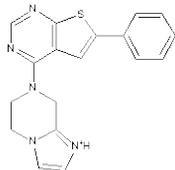 | 333 |
| 11 | 89701933 | ENAMINE Z1170089564 | 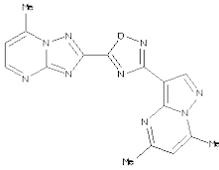 | 347 |
| 12 | 68640014 | ENAMINE Z913154298 | 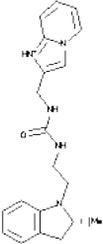 | 350 |
| 13 | 72278988 | ENAMINE Z985662646 | 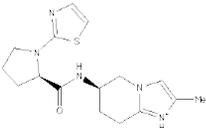 | 332 |

| 14 | 96231403 | InterBioScreen STOCK7S-46707 | 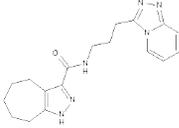 | 338 |
| 15 | 96119804 | Life Chemicals F6413-0713 | 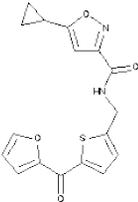 | 342 |